Date: Thu, 11 Feb 1999 14:59:55 -0800 (PST)
From: tom broadhurst <tjb@wibble.Berkeley.EDU>
To: bouwens@wibble.Berkeley.EDU

\documentstyle[11pt,emulateapj]{article}

\begin{document}

\title{Detection of Evolved High-Redshift Galaxies in Deep NICMOS/VLT Images}

\author{Narciso Ben\'\i tez, Tom Broadhurst, Rychard Bouwens}
\affil{Astronomy Department, 
    University of California,  
    Berkeley, CA 94720} 
\author {Joseph Silk} 
\affil{Astronomy and Physics Departments, and Center for Particle 
    Astrophysics, University of California, 
     Berkeley, CA 94720} 
\centerline \& 
\author{Piero Rosati} 
\affil{European Southern Observatory, D-85748, Garching, Germany} 

\begin{abstract}

 A substantial population of high redshift early-type galaxies is
detected in very deep UBVRIJHK images towards the HDF-South.  Four
elliptical profile galaxies are identified in the redshift range
1$<$z$<$2, with very red SEDs, implying ages of
$\gtrsim\hspace{-.15cm}2$ Gyrs for standard passive evolution. We also
find later type IR-luminous galaxies at similarly high redshift, (10
objects with z$>$1, H$<$25), with weak UV emission implying single
burst ages of $\gtrsim\hspace{-.15cm}1$ Gyr. The number and
luminosity-densities of these galaxies are comparable with the local
E/SO-Sbc populations for $\Omega_m$\hspace{-.065cm}$>$0.2, suggesting
that the major fraction of luminous Hubble-sequence galaxies have
evolved little since $z\sim2$. A highly complete photometric redshift
distribution is constructed to H=25 (69 galaxies) showing a broad
spread of redshift, peaking at z$\sim$1.5, in reasonable agreement
with some analyses of the HDF. Four `dropout' galaxies are detected at
$z\approx 3.8$, which are compact in the IR, $\sim$0.5 kpc/h at rest
3500\AA.  No example of a blue IR luminous elliptical is found,
restricting the star-formation epoch of ellipticals to z$\geq$10 for a
standard IMF and modest extinction.

\end{abstract} 
\keywords{galaxies: evolution} 
 
\section{Introduction} 
 
 The detection of early type galaxies at z$>$1 requires deep IR
imaging to overcome the observed large restframe optical K-correction
and correspondingly minimal inferred passive spectral evolution for
this class of galaxy (e.g. Stanford, Eisenhardt \& Dickinson
1998). The highest redshift example of an early-type galaxy is a
luminous object reported by Dunlop et al.\ (1996) at z=1.55. Its
restframe UV spectrum is dominated by late-type F-stars, equivalent to
an age of $\approx 3$Gyrs, for standard stellar synthesis models
(Spinrad et al.\ 1997). Following ordinary ellipticals to $z>1$
requires much deeper IR imaging now possible from space. The NICMOS 
observation in the Hubble Deep Field South (Fruchter et al 1999)
represents a three magnitude increase in depth over the deepest ground
based IR data, and with usefully high spatial resolution. This, in
combination with deep optical images from the VLT allows easy
identification of red $L{^*}$ ellipticals to $z\sim2.5$, where
significant spectral and density evolution is widely anticipated.

In section $\S1$ we briefly outline our spectral and profile fitting
procedures and simulations as applied to the NICMOS and VLT observations.  
In $\S2$ we discuss detections of early-type and other IR bright galaxies
at high--redshift. In $\S3$ we compare their statistical properties
with local populations of Hubble sequence galaxies of similar spectral
type, for different evolutionary models.  In $\S4$ we briefly discuss a
the absence of bright blue precursor
ellipticals in the IR.

\section{Photometry, Redshift Estimation and Profile Fitting}

Three sets of images are utilized, the HDF-S NICMOS observations in
JHK (Fruchter et al. 1999), the higher resolution unfiltered HST STIS
image (Gardner et al. 1999) and the VLT Test Camera images in UBVRI
(Fontana et al. 1999). We use the NICMOS pipeline zero-points and have
computed the VLT effective zero-points for the combined optical images
from the ESO photometric solution. The filter functions and instrument
response curves are kindly made available by ESO and STScI. The
detection is performed using SExtractor (Bertin \& Arnauts 1996) in a
combined $J+H$ frame. Near total magnitudes are measured in J,H,K and
PSF matched STIS images within a matched aperture set. A total of
$\approx 200$ galaxies are detected to H=27 within the central
50''$\times$ 50'' field. The VLT UBVRI images have PSF's $\approx 3$
times wider that the NICMOS exposures, and require a different
procedure. We measure magnitudes within a $0\farcs 9$ aperture and
correct them to near-total magnitudes using versions of the $J+H$ band
image degraded to the PSF of each of the VLT passbands to ensure the
colors are defined within the same apertures for all bands.

Photometric redshifts are estimated using the methods described 
in Ben\'\i tez (1999). The empirical template set of Coleman, Wu, \&
Weedman (1980) is used, and augmented with bluer local starbursts
(Kinney et al.\ 1996), to determine the redshift probability 
distribution for each galaxy with Bayesian weighting.  
The fraction of objects with highly reliable photometric redshifts
(p$>$0.99) is $57/69$ to H$<$25 ($24/25$ for objects classified as
earlier than Scd).  Fainter than this magnitude limit the reliability
of the photometric redshift estimates quickly degrades.  
We also allow for the quoted 10\% uncertainty in the
NICMOS zero-points for our spectral fits.

For redder early types, more detailed fits to single burst synthetic
models are justified to accommodate the known spread of metallicity and
age. For this, the compilation of Leitherer et al.\ (1996) is used,
including the synthetic galaxy spectra of Bruzual \& Charlot (1993).
We also check that the colors of these objects are not compatible with
dust reddened star--forming galaxies by generating reddened versions
of the late and Im types in our template set. 

Profile fitting is performed for bright galaxies in the IR (H$<$23.7).
For this we simulate NICMOS images using the methods described in
Bouwens, Broadhurst \& Silk (1998) to account for the pixelization,
dithering and point spread functions and the available weight map for
the NICMOS dithering pattern.

\section{High-Redshift Red Galaxies}

The $H<25$ sample contains five clear elliptical galaxies, of which four lie above
z=1. Their SED's and surface brightness profiles are compared with the
models in Fig~1 and listed in Table~1. All the objects spectrally
classified as ellipticals are also found to have spheroidal profiles
(Fig~1). The most impressive object (NIC3/ET4, Fig~1, Table~1) is extremely
red, consistent with minimal passive evolution at an estimated
$z=1.94\pm0.15$, with a clear de-Vaucouleurs profile and uniform color
over its surface ($\sim$ 300 pixels).  The effective radius
1.6(2.2)kpc/h is consistent with its inferred luminosity
$M_{AB}=-21.96(-22.6)$ in restframe R for $\Omega_m=1(0.1)$
(Binggelli, Sandage \& Tarenghi 1984).  This IR conspicuous object is
absent or only marginally detected in the U,B,V,R,I and STIS bands
(Fig~1). 
Dust reddened spectra extinguished with standard reddening
(Calzetti, Kinney, \& Storchi-Bergmann 1994) are unable to generate
the steep break from the relative flat IR.  
Another elliptical identified at $z=1.66 \pm 0.25$ (NIC3/ET3, Fig 1, Table1) with a 
clear de-Vaucouleurs profile, is the luminous object reported by 
Treu et al. (1998) in the earlier NICMOS test images and confirmed by 
Stiavelli et al. (1999).

We also find several IR-bright high-redshift galaxies of later 
spectral type, 10 objects with z$>$1 and H$<$25.  Two typical examples
are shown in Fig~1.  The profiles of these objects are generally
exponential.  The SED's are very conspicuous with a turn up towards
the UV, and a distinctive bump in the IR separated by a trough,
so there is little uncertainty in redshift (e.g. Fig~1). The best
fit single burst minimum ages are typically $\sim 1$ Gyr. Comparison
with the SEDs of CWW show these objects span the spectral types of
local Sab to Sbc galaxies. The full sample of 24 galaxies
with H$<$25 (open histogram, Fig~2) have a redshift
distribution peaked at $z~1.5$ with high-z tail similar to that found
for the HDF (Connolly et al.\ 1997, Fern\'andez--Soto, Lanzetta \&
Yahil 1998, Ben\'\i tez 1999).

\section{Density Evolution}

A reliable measurement of the space density of massive galaxies at
high redshift provides a basic check of hierarchical models of galaxy
formation, where the most massive haloes form late (Lacey \& Cole
1993) and in which ellipticals and massive spheroids may be viewed as
the end product of the merging of smaller systems.

The presence of so many luminous red galaxies at high-z in such a
small field is surprising and indicates that relatively evolved
luminous galaxies are commonplace at z$>$1.  Using the Pozzetti,
Bruzual, \& Zamorani (1996) local luminosity functions for E/SO-Sbc as
input, we recover 4(9) ellipticals from our realistic NICMOS
simulations for $z>1$, $H<25$, with $\Omega_m=1(0.2)$ and no
evolution.  Adding the later-type IR-luminous galaxies allows a useful
comparison with models (Fig~2).  The redshifts and luminosities
observed above z=1 are consistent with no density or luminosity evolution
of standard Hubble sequence galaxies for $0.2<\Omega_m<1$.  Note, passive
evolution is minimized here for consistency with the red SEDs
by choosing $H_0=50$km/s/Mpc and $z_f=15$. The hierarchical model
parametrization of Kauffmann, Charlot \& White (1996) for $\Omega=1$
underpredicts the numbers as expected. Hierarchical models with $\Omega <1$ 
may allow a declining density but must be checked for self consistency. 
The optical-IR colors prefer older higher redshift models
and are consistent with no size evolution, but do not distinguish
clearly between the above cosmological models.

The integrated luminosity density to H=25 at 1$<$z$<$2.5 is comparable
with local estimates integrated over E/SO-Sbc types (see Fig~2). The
predicted luminosity density for $z>1$ at restframe 6000$\AA$ is
$4.3\times10^{19}$ergs/cm$^{2}$/s/Hz compared to $3.9\pm1.4\times
10^{19}$ ergs/cm$^{2}$/s/Hz, $2.41\pm0.9\times 10^{19}$
ergs/cm$^{2}$/s/Hz and $2.06\pm0.8\times 10^{19}$ ergs/cm$^{2}$/s/Hz,
for $\Omega_m=1$, $\Omega_m=0.1$ and
$\Omega_m+\Omega_{\Lambda}=0.25+0.75$ respectively.  A factor of two
decrease in luminosity density is implied relative to the popular low
$\Omega_m$ models assuming passive or no stellar evolution.

Our results seem to extend to $z\approx 2$ the results of other authors 
notably Lilly et al. (1995) and Kodama, Bower \& Bell (1999) who find no 
evidence for a dramatic evolution in the density of early types galaxies to 
$z\sim1$. The contrast of our results with the absence of distant ellipticals
claimed by others (Kauffmann \& Charlot 1998; Zepf 1997; Franceschini
et al.\ 1996; Barger et al.\ 1998; Menanteau et al.\ 1998) can be put
in context by appreciating that the NICMOS data is complete to
$H\approx 27$, which is 3 magnitudes fainter than the deepest ground
based data and 6 magnitudes fainter than most work on this
question. Two of the four ellipticals found here in the IR would
barely register in the I band (I$\sim$27) and are certainly not bright
enough for morphological classification in the optical even from space. 
Recent claims of a decline in the elliptical density for z$>$1 rely 
on significant passive evolution, where brightening enhances the 
expected numbers.
The claim by Kauffmann, Charlot, \& White (1996), based on the CFRS
(Lilly et al. 1995) is subject to redshift incompleteness, in
particular of red high-z galaxies, which are very hard to measure
spectroscopically. Some of the red galaxies
reported here are bright enough in the IR to be detected in previous
deep Keck images (Moustakas et al. 1997; Hogg et al. 1997) but
morphological classification is not possible in ground based images
and deep complementary optical data are required to establish the break.
The NICMOS images directly show that luminous early type galaxies are
present in significant numbers at high redshift with SEDs too red to
be recognized as ellipticals in the optical even in the deepest WFPC2
images.

An obvious problem with this kind of work is the presence of
clustering in small field, rendering the results uncertain.  To
explore this we have searched for spikes in the redshift distribution
using the approach described in Ben\'\i tez (1999).  The
minimal number of structures along the light--of-sight to H=25 is
three, at $z=0.55$ (consistent with the $0.58$ redshift spike found in
the HDF-S by Glazebrook et al. 1999), $z=1.35$ and $z=1.95$. 
Each of these structures corresponds approximately to the redshift of 
one or more elliptical
galaxies, and the two higher redshift peaks contain $\approx 1/3$ of all
the $z>1, H<25$ galaxies. Our statistical analysis shows
that it is virtually impossible that all the early type galaxies at
$1<z\lesssim 2$ belong to a single spike.  A rough estimate of the
expected clustering noise can be obtained by 
integrating the faint galaxy correlation function over the NICMOS 
field for the range 1$<$z$<$2. Connolly, Szalay \& Brunner 1998 
measure an amplitude $A(10'')=0.12 (\omega \propto \theta^{-0.8})$ for 
the angular correlation of HDF-N galaxies within $\Delta z=0.4$ slices to $z\sim 1.5$. 
Assuming that $A\propto 1/\Delta z$, the total variance 
in the number of galaxies in a redshift slice $\Delta z$ is 
$\sigma^2(\Delta z)\sim N(\Delta z)[1+N(\Delta z)(0.4/\Delta z)0.07]$.
For the interval $1<z<2$ with 35 galaxies this yields a variance of
$\sigma=8.3$ galaxies, which is 40\% greater than Poisson. 
If the early types are clustered like the integral of all types
then we estimate their number to be $14\pm4$ in the range $1<z<2$.

\section{Very High-Redshift Blue Galaxies}

The lack of spectral evolution of early type galaxies demonstrates
that star-formation does not extend over time, but is confined to an
early brief period, so that we should expect the precursors of these
galaxies to appear extremely luminous and blue at higher redshift. The
width of this period is constrained to be short by the tight E/SO
color-magnitude and M/L sequences (e.g.  Bower, Lucey \& Ellis 1992,
VanDokkum et al.\ 1998).  More directly, the lack of blue ellipticals
in the HDF (e.g. Franceschini et al.\ 1998) means this epoch lies at
$z>5$, and therefore may appear in the near-IR for redshifts in range
5$<$z$<$15 as I or J band `dropouts', provided the IMF of early-type
galaxies is not unusually deficient in hot stars.

Our simulated NICMOS images show that significant numbers of
elliptical galaxies are expected in the IR if they form luminous stars
high redshift and were not significantly extinguished during the first
Gyr. The simulations include an exponential burst at $z_f=15$ with an
e-folding time of 1Gyr and predicts 6, 10 and 15 objects for z$>$5 to
H=26 for, $\Omega_m=1$, $\Omega=0.1$,
$\Omega_m+\Omega_{\Lambda}=0.25+0.75$. In comparison, no I or J band
`dropout' galaxies are observed brighter than this limit.

At lower redshift, five compact optically blue dropout galaxies are
found, all consistent with a single redshift of $z\approx 3.8$. Fits
to their spectra are shown in Fig~3. These objects have low SFR of
6-23$M_{\odot}/yr$ and very compact sizes in the H-band, $\sim$0.5
kpc/h at restframe $3500\AA$ ($\Omega_m$=1), and therefore appear
unlikely to be the precursors of luminous elliptical galaxies if no
evidence of SF is to remain by z$\sim$1.

\section{Conclusions}

We have detected a substantial population of early type galaxies at
z$>$1. These include $r^{1/4}$ profile galaxies in the redshift range
z$>$1, all of which have surprisingly red colors despite their large
redshifts. Later spectral-types are also present in this redshift
range, with a conspicuous IR bump. These galaxies are all luminous and
well evolved, with minimum single-burst ages $>$1Gyr for a standard
IMF.

The clear detection of early type galaxies and spheroidal components
of other high-z disk galaxies points to the existence of a roughly
similar density of luminous Hubble sequence galaxies at $z\sim2$,
compared to the local Universe. The declining density of luminous
galaxies expected in hierarchical schemes can be made consistent with
the data only with for low $\Omega_m$ models. Low $\Omega_m$ helps
accommodate the absence of spectral evolution, but in any cosmology the
lack of bright IR-blue precursors of elliptical galaxies requires a
very high redshift of formation z$>$10, for a standard IMF and
extinction.

Although the field studied is small, the number of red high-z
galaxies detected is surprising and strongly suggests that the bulk of
luminous galaxies were formed at $z\gtrsim2$. In the short--term a deep
NICMOS pointing in the HDF will provide similar data and help average
over clustering. Progress on this question awaits
deep diffraction-limited IR imaging on large telescopes and improved
IR facilities in space.

\acknowledgements

We would like to thank Bob Williams and the Hubble Deep Field South
Team as well as Alvio Renzini and the VLT-UT1 Science Verification
Team for making available this useful data.

\begin{deluxetable}{cccccccc}
\tablenum{1}
\tablewidth{0pt}
\tablecaption{$z>1$ Early type galaxies in the HDF/NIC3 field }
\tablehead{ name & \colhead{X} &
\colhead{Y} & $m_{F160W,AB}$ & $z_{phot}$ & 
$r_{hl}$ & $STIS_{AB} - F160W_{AB}$ &
\colhead{Age(Gyr)}}
\startdata
NIC3/ET1 & 181.6 & 868.87 & $21.24\pm0.01$ & $1.41\pm0.20$ & $0\farcs34$ & 3.3 & $>1.5$  \nl
NIC3/ET2 & 439.2 & 352.34 & $24.50\pm0.06$ & $1.55\pm0.20$ & $0\farcs31$ & $>4.0$ & $>1.5$ \nl
NIC3/ET3 & 607.4 & 371.52 & $21.79\pm0.01$ & $1.66\pm0.25$ & $0\farcs26$ & 4.2 & $>2$ \nl
NIC3/ET4 & 646.8 & 898.32 & $23.18\pm0.02$ & $1.94\pm0.15$ & $0\farcs39$ & $>5.4$ & $>3.5$ \nl
\enddata
\end{deluxetable}

\begin{figure}[t] 
\epsscale{1}
\plotone{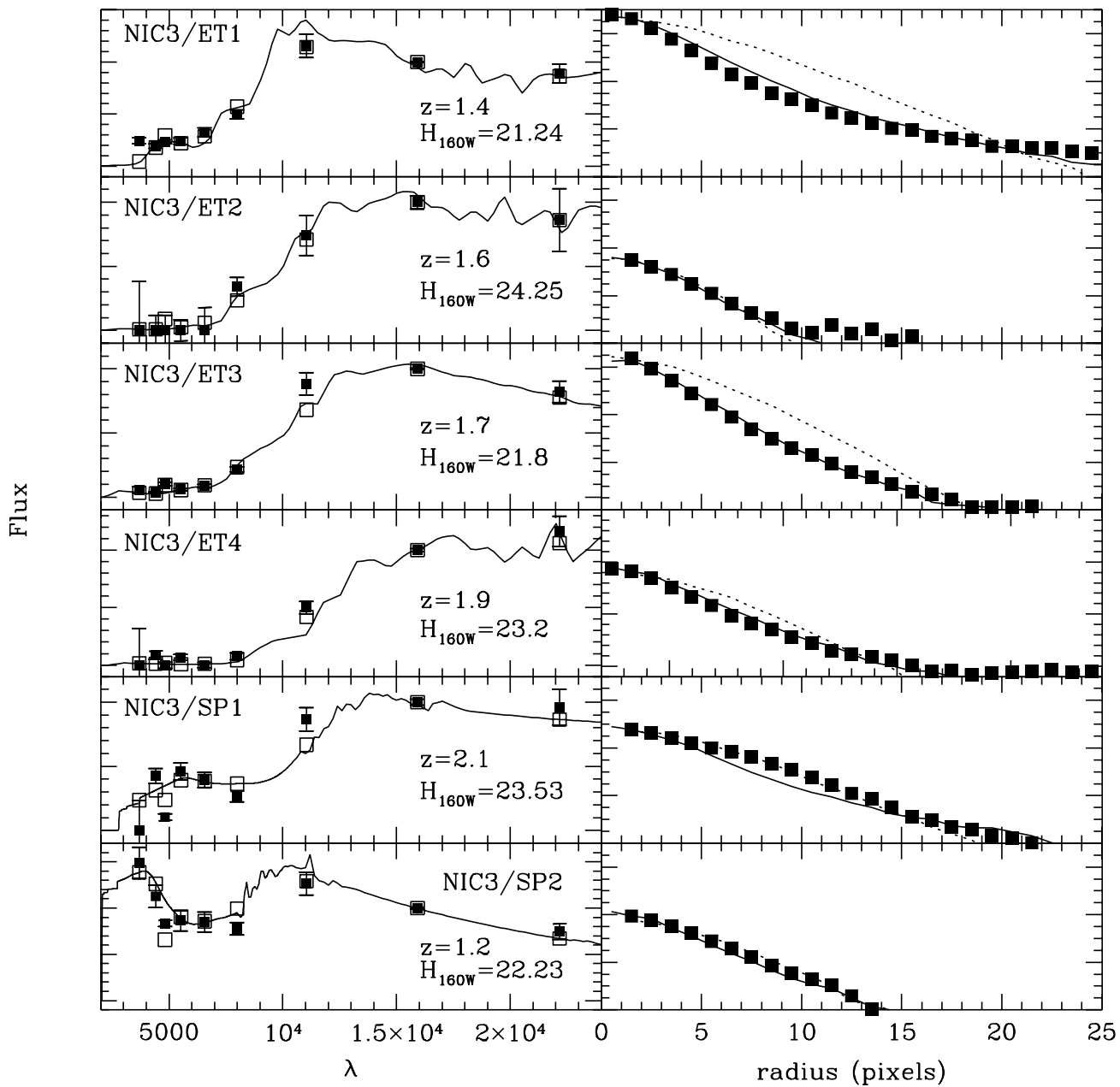}
\caption{The left panel shows our best fit elliptical galaxies at
z$>$1.  On the left is the SED fit to the multiband photometry where
the data points should be compared with the integrated flux over the
passband from the best fit model and represented as open squares. The
corresponding model spectrum is also shown. On the right are the
convolved exponential (dotted) and de-Vaucouleurs (solid) best fits to
the data, plotted as log surface-brightness versus linear radius for
clarity. The two lower panels are representative examples of later
spectral types at high redshift (Sab and Sbc), demonstrating the
utility of U through K broadband information for identifying redshifts
of these galaxy types at high redshifts. The positions of objects ET1 
through ET4 are listed on Table1; objects SP1 and SP2 have coordinates 
 (595.8, 1001.3) and (835.4, 1022.2).}
\end{figure}

\begin{figure}[t] 
\epsscale{1}
\plotone{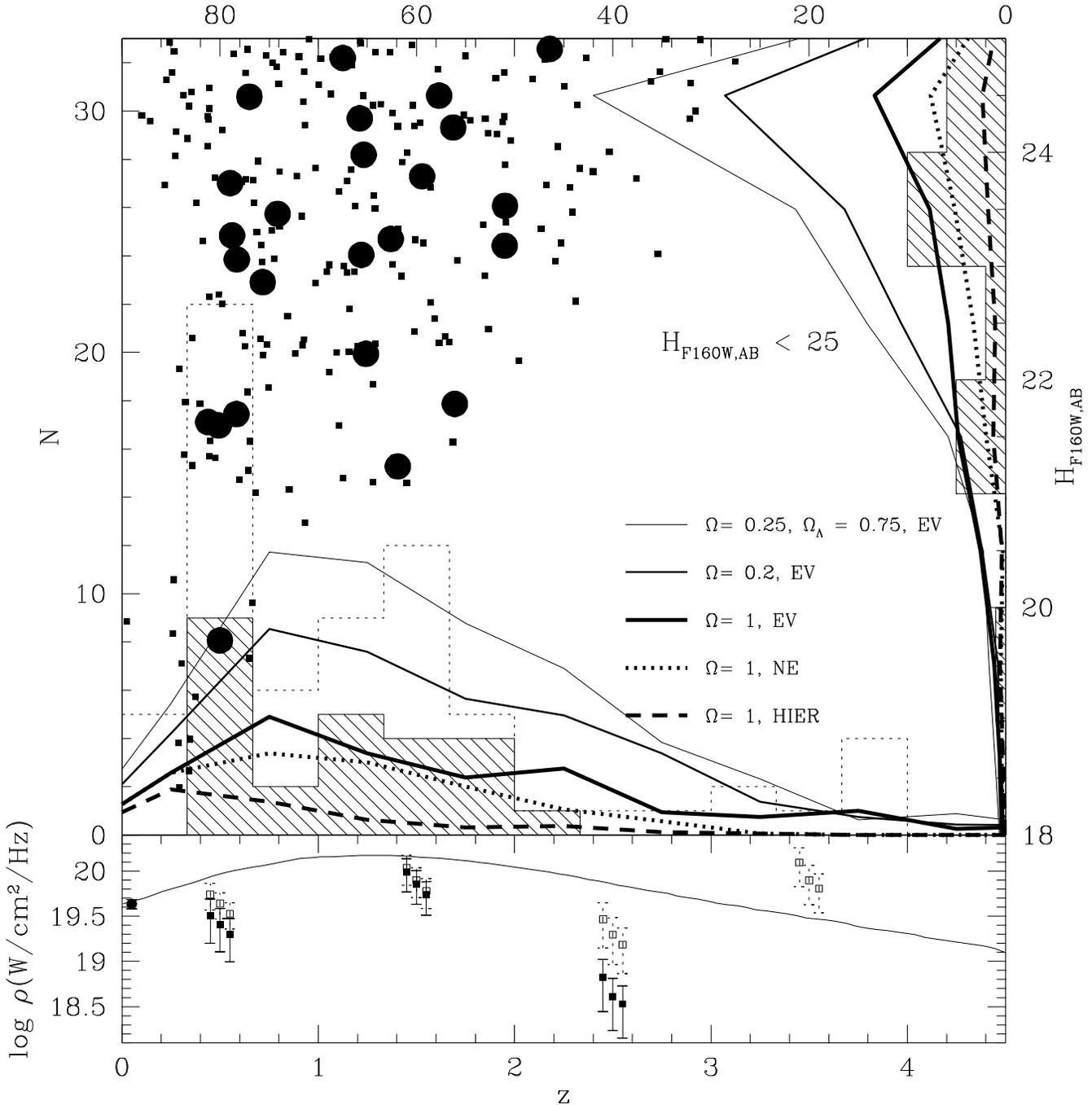}
\caption{Models of local E/SO-Sbc luminosity functions recovered from
simulated images matched to the properties of the NICMOS South data.
These are compared with the redshift distribution of E/SO-Sbc spectral
types to $H=25$ in the lower histogram and against apparent magnitude
to the left.  The predicted range of redshifts is similar to the data
(shaded histogram) for all models, but the observed numbers are
greater than expectations of hierarchical models for massive galaxies
with $\Omega_m=1$ and underpredict no-evolution or mild passive
evolution with larger volume cosmologies.  Consistency is found with
mild or no stellar evolution for relatively flat $\Omega_m$ dominated
models.  Several Monte-Carlo realizations of the redshift-magnitude
distribution for a no-evolution $\Omega=1$ model (small points) are
compared to the observations (big points).  Dotted histograms give the
redshift distribution for all types to $H=25$.  The lower panel shows
the luminosity density with redshift for the spectrally early-type
galaxies (solid squares) and for the whole sample (open squares) to
$H=25$ at rest $6000\AA$.  Compared with the local luminosity-density
of E/SO-Sbc galaxies (left most point), there is no obvious trend.
The curve is our calculation of the luminosity density at rest
$6000\AA$ using the Madau, Pozetti \& Dickinson (1998) SF history for
the Salpeter IMF and no dust.}
\end{figure}

\begin{figure}[t] 
\epsscale{1}
\plotone{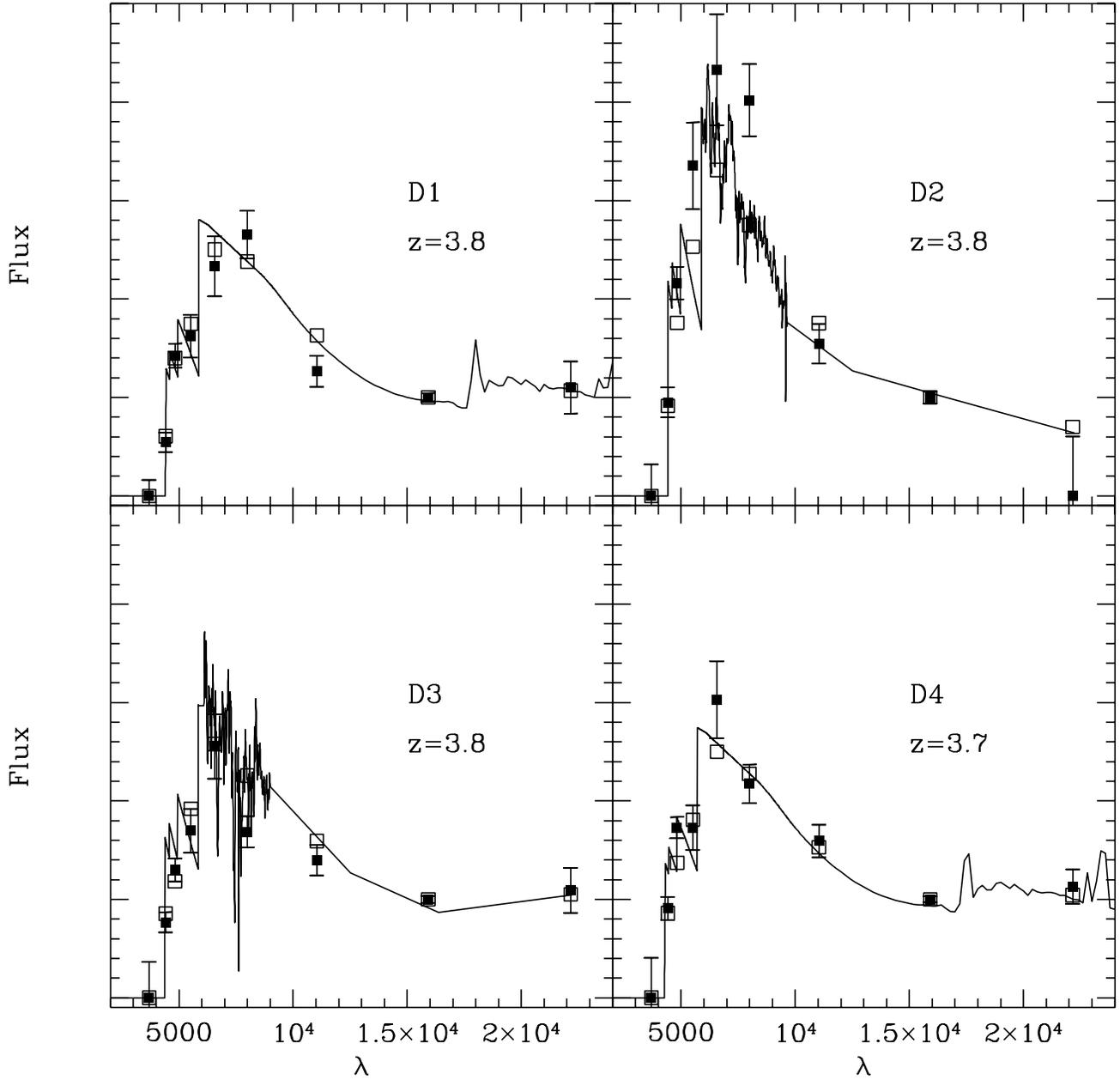}
\caption{The optical dropout galaxies showing the best single burst
minimum ages ranging from $10^6$-$10^8$ yrs and star formation rates
of between 6-23$M_{\odot}/$yr for a Salpeter IMF and $\Omega=1$. These
blue objects are very compact in the IR $\sim$0.5 kpc/h at rest $3500\AA$
and all consistent with lying at $z\approx3.8$. No example of a
higher redshift IR-dropout galaxy up to $H<26$ is found, placing 
a useful constraint on single burst models of luminous galaxy 
formation (see text). D1,D2,D3 and D4 have respectively positions 
(244.3, 598.1), (318.3, 855.7), (857.6, 847.4) and (497.8, 770.7)}
\end{figure}

\end{document}